\documentstyle[epsfig,aps]{revtex}
\begin{document}
\draft
\title{Motional rotating wave approximation for harmonically trapped particles}
\author{\"Ozg\"ur E. M\"ustecapl{\i}o\u{g}lu and L. You}
\address{School of Physics, Georgia Institute of Technology,
Atlanta, GA 30332-0430, USA}
\date{\today}
\maketitle
\begin{abstract}
We present a family of generalized unitary transformations that
simplifies the Hamiltonian for a harmonically trapped two level atom
(or ion) interacting with a plane wave laser field. Novel near
resonant single as well as double vibrational phonon dynamical
regimes are found. The validity condition of the often used motional
rotating wave approximation (MRWA) is examined both numerically
and analytically. Large errors are found within typical regimes of
MRWA with respect to the motional degrees of freedom. The effects
of MRWA in trapped ion systems are shown to be opposite to that of
the rotating wave approximation (RWA) in the usual Jaynes-Cummings
model. Our study points to a more restrictive condition on
particle localization (Lamb-Dicke) parameter for the validity of
MRWA in the single phonon dynamical regime. It also sheds new
light on quantum information storage and processing with trapped
atoms.
\end{abstract}



\section{Introduction}
\label{sec:intro}
In the last few years, much attention has been focused on quantum
dynamics and coherence properties of trapped atoms or ions
\cite{fang,jonathan,moya,wu,mancini}. These studies led to many
potentially attractive applications of harmonically trapped
particles, e.g. in generating nonclassical vibrational phonons
\cite{cirac}, implementing fast quantum gates \cite{jonathan}, and
in achieving phonon-ion entanglement \cite{fang}. In most of these
studies, a coherent plane wave laser field near-resonantly couples
two electronic states of an atom. The inclusion of motional degrees of freedom
leads to tremendous complication. Several simplifications have been
developed that reduce a trapped particle dynamics to the
familiar Jaynes-Cummings Model (JCM) type form with the use
of Lamb-Dicke limit (LDL), the strong confinement limit, or
the motional rotating wave approximation (MRWA). In fact, the JCM
description can be achieved regardless of laser field configuration,
i.e. whether it is a travelling-wave \cite{moya} or a standing-wave
\cite{fang,wu}. LDL requires the particle localization size,
given by the harmonic trap ground state width $a$, to be much less
than the near resonant laser wavelength $\lambda$, i.e.
$\eta=(2\pi)a/\lambda\ll 1$. It is often believed that
MRWA for harmonically trapped particles works well within
LDL as it demands a less restrictive condition $\eta \ll
2$\cite{moya} or $\eta \ll 4$\cite{jonathan}.

In our ongoing effort to understand motional effects of trapped
particles for quantum information processing \cite{you}, this
issue of validity regimes for trapped particle MRWA again arises.
We note some earlier investigations show that rotating wave approximation
(RWA) does not work satisfactorily for Hamiltonians containing multiple
transitions \cite{tur,fang2} or for systems initially prepared in a
superposition of internal states \cite{seke}. Furthermore,
multi-particle properties such as entanglement can have different
sensitivity dependence on single particle MRWA \cite{xie}.
With harmonically trapped particles, this issue becomes
particularly acute as the equal distant motional states span an
infinite dimensional Hilbert space. MRWA is typically made after
a linearization by either taking LDL or by applying an exact
simplifying unitary transformation. Within the latter approach,
errors from MRWA can be further modified by the back transformation
when dynamical observables are calculated in the original frame.

In this article we study quantum dynamics of a single trapped atom
by employing a numerical diagonalization procedure without
employing MRWA. We assess the validity regime for MRWA by
comparing with results from analytical models obtained under MRWA.
This paper is organized as follows. In Sec. II, our model system
of a single trapped two level atom is described and a review of
the unitary transformation method for linearizing the Hamiltonian
is presented. The new result, the existence of a
family of general unitary transformations for linearizing our model
Hamiltonian, is then introduced. In Sec. III, we discuss two
linearized models obtainable from our transformations. In Section
IV, we outline several technical points about analytical
solutions of the transformed model Hamiltonian under MRWA. We also
discuss a numerical diagonalization procedure used for
exact dynamic solutions. Selected results and comparisons are
presented in Sec. V. Finally we conclude in Sec. VI.
\section{The model system and a family of general unitary transformations}
\label{sec:utm}
To simplify our discussion, we consider a one-dimensional model of
a harmonically trapped two level particle interacting with a near
resonant laser field \cite{moya,lewenstein}. The system Hamiltonian
is given by
\begin{eqnarray}
{\cal H}&=&{\cal H}_0+{\cal H}_1, \nonumber\\
{\cal H}_0&=&\frac{P^2}{2M}+V_{tg}(x)\sigma_{gg}
+[\hbar\omega_{eg}+V_{te}(x)]\sigma_{ee},\nonumber\\
{\cal
H}_1&=&\frac{\Omega}{2}e^{i\hbar\omega_Lt}e^{-ik_Lx}\sigma_-+{\rm
H.c.} \label{h01}
\end{eqnarray}
where $\Omega$, $\omega_L$, and $k_L$ denote respectively the
Rabi frequency, carrier frequency, and wave number of a
coherent driving laser.
The electron transition frequency between excited ($|e\rangle$) and
ground state ($|g\rangle$) is $\omega_{eg}$. $\sigma_{ab}=|a\rangle\langle b|$
($a/b=e,g$) are atomic projection operators.
Consist with convention we denote $\sigma_-=|g\rangle\langle e|$ and
$\sigma_z=\sigma_{ee}-\sigma_{gg}$. For a neutral particle,
typically the approximate harmonic trap potential is internal state
dependent, i.e. $V_{ta}=(1/2)M\nu_a^2x^2$ with a corresponding trap
frequency $\nu_a$. $P$ is the motional momenta and $M$ is the mass
of the particle.

We first simplify Eq. (\ref{h01}) by changing into the interaction
picture (rotating frame) with the unitary transformation ${\cal
H}\rightarrow e^{i\hbar\omega_Lt\,\sigma_{ee}}{\cal
H}e^{-i\hbar\omega_Lt\,\sigma_{ee}}-\hbar\omega_L\sigma_{ee}$.
This leads to
\begin{eqnarray}
{\cal H}_0&=&\frac{P^2}{2M}+V_{tg}(x)+\frac{\hbar\delta}{2}\sigma_{z}+
[V_{te}(x)-V_{tg}(x)]\sigma_{ee}+\frac{\hbar\delta}{2},\nonumber\\
{\cal H}_1&=&\frac{\Omega}{2}e^{-ik_Lx}\sigma_-+{\rm H.c.},
\label{h02}
\end{eqnarray}
where $\delta=\omega_{eg}-\omega_L$ is the laser field detuning.
We now introduce motional phonon annihilation (creation) operator
for the ground state $a$ ($a^{\dagger}$) according to
$P=\sqrt{\hbar M\nu_g/2}\,p$ and $x=\sqrt{\hbar/2M\nu_g}\,q$ with
$p=i(a^{\dagger}-a)$ and $q=a^{\dagger}+a$. The displacement
operator becomes $D(\beta)=e^{\beta a^{\dagger}-\beta^{\ast}a}$,
and Eq. (\ref{h02}) simplifies to $(\hbar=1)$
\begin{eqnarray}
{\cal H}_0&=&\nu_gn+\frac{\delta}{2}\sigma_z+\zeta\,q^2\sigma_{ee}
+\frac{\nu_g+\delta}{2},\nonumber\\
{\cal H}_1&=&\frac{\Omega}{2}D^{\dagger}(i\eta)\sigma_-+{\rm H.c.},
\end{eqnarray}
where $n=a^{\dagger}a$ is the phonon number operator and
$\zeta=(\nu_e^2-\nu_g^2)/4\nu_g$. The Lamb-Dicke parameter is now
$\eta=k_L\sqrt{1/2M\nu_g}$. For this particular form of the
Hamiltonian, we note that its interaction part (${\cal H}_1$) can
be diagonalized by a general transformation matrix
$\tilde{T}=E^{\dagger}T$, with
\begin{eqnarray}
E&=&\frac{E_2+E_1}{2}\hat{I}+\frac{E_2-E_1}{2}\sigma_z,\nonumber\\
T&=& \frac{1}{\sqrt{2}}\left(\frac{D^{\dagger}(i\eta)+1}{2}\hat{I}
+\frac{D^{\dagger}(i\eta)-1}{2}\sigma_z+\sigma_+-D^{\dagger}(i\eta)\sigma_-\right).
\end{eqnarray}
We choose $\Omega \in \Re$ without loss of generality. It is easy
to check that $\tilde{T}{\cal H}_1\tilde{T}^\dag
=(\Omega/2)\sigma_z$, is independent of the arbitrary unitary
functional operators $E_{1,2}$ of $a$ and $a^\dag$. For any
operator $\Theta$, the transformed operator will be denoted to be
$\tilde{\Theta}=\tilde{T}\Theta\tilde{T}^\dag$.
$\hat{I}=\sigma_{ee}+\sigma_{gg}$ is the identity operator. This
transformation can be compared to the generalized Power-Zienau
transformation discussed earlier \cite{drummond}. In the phonon
Fock state basis, it generates coherent superpositions of motional
wave-packet states which was previously used in studying motional
decoherence of atomic qubit operations \cite{you}. It reduces to a
simplifying transformation used by Moya-Cessa {\it et al.}
\cite{moya} when we take $E_1=E_2=D(-i\eta/2)$. The $\tilde{T}$
transformation on the ${\cal H}_0$ term can be conveniently
calculated using the following properties
\begin{eqnarray}\label{eq:prop1}
D(\alpha)D(\beta)&=&e^{(\alpha\beta^{\ast}-
\alpha^{\ast}\beta)/2}D(\alpha+\beta),
\label{t1}\\
D(\mp i\eta)a^\dag a D(\pm i\eta)&=&a^\dag a\pm i\eta(a^\dag -a)+\eta^2,
\label{t2}
\end{eqnarray}
for arbitrary complex numbers $\alpha$ and $\beta$. We note that
Eq. (\ref{t1}) simplifies to $D(\alpha)D(\beta)=D(\alpha+\beta)$
when $\alpha$ and $\beta$ are purely imaginary. The second and
the last term in the rhs of Eq. (\ref{t2}) correspond
to the Doppler and recoil shifts. After this general
transformation, we obtain
\begin{eqnarray}
\tilde{\cal H}_0 =E^\dag \left[(\nu_gn+gp+\eta
g+\frac{\delta+\nu_g}{2}+\frac{\zeta}{2}q^2)\hat{I}
-(gp+\frac{\zeta}{2}q^2+\eta g+\frac{\delta}{2})\sigma_x\right]E,
\label{th0}
\end{eqnarray}
with $g=\eta\nu_g/2$ and $\sigma_x=\sigma_++\sigma_-$. Different
simplifications can be pursued by exploiting other forms of
$E_{1,2}$. We note that the first term in Eq. (\ref{th0}) is in
the form of a squeezed displaced harmonic oscillator. Therefore it
can be diagonalized in the Fock state basis by the squeezed
coherent state transformation $E_2=E_1=D(\beta)S(\xi)$. To
eliminate the $p$ dependence in the coefficient of $\hat{I}$ we
choose $\beta=-i\eta/2$, which transforms according to
$n\rightarrow n-(\eta/2)p+\eta^2/4$, $p\rightarrow p-\eta$, and
$q\rightarrow q$. We then obtain
\begin{eqnarray}\label{eq:s-h0}
\tilde{\cal H}_0 =S^{\dagger}(\xi)
\left[\left(\nu_gn+\frac{\zeta}{2}q^2+\frac{\delta+\nu_g+\eta g
}{2}\right)\hat{I}
-\left(gp+\frac{\zeta}{2}q^2+\frac{\delta}{2}\right)\sigma_x\right]S(\xi).
\end{eqnarray}
The squeezing transformation $S(\xi=re^{i\theta})$ causes
$a\rightarrow ua^{\dagger}-v^{\ast}a$ with
$u=\cosh{r}$ and $v=e^{i\theta}\sinh{r}$.
For our model, $\zeta$ is required to be real,
i.e. $\theta=0$. The action of $S$ results in
\begin{eqnarray}
p&&\rightarrow e^rp,\nonumber\\
q&&\rightarrow e^{-r}q, \nonumber\\
n&&\rightarrow e^{2r}n-\frac{1}{2}\sinh{2r}q^2+e^r\sinh{r}.
\end{eqnarray}
Substitution these results into Eq. (\ref{eq:s-h0}) leads to the
elimination of the $q^2$ term in the coefficient of $\hat{I}$ term
provided we choose $r=(\ln{\epsilon})/4$ with
$\epsilon=1+2\zeta/\nu_g$. This yields
\begin{eqnarray}
\tilde{\cal H}_0=\left[\nu_g\sqrt{\epsilon}\,n+
\frac{\nu_g\sqrt{\epsilon}+g\eta+\delta}{2}\right]\hat{I}
-\left(g\epsilon^{1/4}p+\frac{\zeta}{2\sqrt{\epsilon}}\,q^2
+\frac{\delta}{2}\right)\sigma_x.
\end{eqnarray}
We can now redefine parameters according to
$\nu=\nu_g\sqrt{\epsilon}$, $g\rightarrow g\epsilon^{1/4}$, and
$\zeta\rightarrow\zeta/\sqrt{\epsilon}$. Finally we arrive at the
transformed Hamiltonian
\begin{eqnarray}\label{eq:h-quad}
\tilde{{\cal H}}=\nu n+\frac{\Omega}{2}\sigma_z
-\left(gp+\frac{\zeta}{2}\,q^2+\frac{\delta}{2}\right)\sigma_x,
\end{eqnarray}
where the constant term in the coefficient of $\hat{I}$ has been
dropped. We note that for ions with internal state independent
trap frequencies $\zeta=r=\xi=0$. We recover the same result as in
Ref. \cite{moya} by choosing $\beta=-i\eta/2$. The effects of
different trap frequencies for different internal levels are a
renormalization of energy parameters in the single vibrational
phonon Hamiltonian and a double vibrational phonon interaction
through a term quadratic in the position operator. Double phonon
transitions are the typical interactions that can lead to squeezed
and entangled states for the vibrational phonons. In the next
section, we will see that such two-phonon transitions can be
dominated the single-phonon transitions in the system, independent
of the Lamb-Dicke parameter $\eta$.
\section{Motional rotating wave approximation}
\label{sec:ndm-rwa}
The simplified model Hamiltonian Eq. (\ref{eq:h-quad}) is of the
form of a generalized JCM involving quadratic two-phonon
transitions. It can be analytically solved for small squeezing
parameters without external drive ($\Omega=0$) \cite{tannoudji}.
Alternatively, when $|\nu-\Omega|\ll\Omega$ or
$|2\nu-\Omega|\ll\Omega$, and $g$ and $\zeta \ll\Omega$, an
approximate analytic solution can be obtained by an explicit
diagonalization upon the elimination of the rapidly oscillating
terms, i.e. the application of MRWA. In the interaction picture,
Eq. (\ref{eq:h-quad}) becomes
\begin{eqnarray}
\tilde{\cal{H}}_I=\left[-ig(a^{\dagger}e^{i\nu t}-ae^{-i\nu
t})+\frac{\zeta}{2}(a^{\dagger 2}e^{i2\nu t} +a^{\dagger}a+h.c.)+
\frac{\delta}{2}\right](\sigma_+e^{i\Omega t} +\sigma_-e^{-i\Omega
t}). \label{eq:ndm}
\end{eqnarray}
There exist two resonance conditions, one at $\Omega=\nu$ where
the validity condition for MRWA becomes
$(\eta\nu_g/2)\epsilon^{1/4}\ll \nu_g\epsilon^{1/2}$
\cite{tannoudji}, i.e. $\eta\ll 2\epsilon^{1/4}$. When the
difference in trap frequencies is small we get $\eta\ll
7/4+(\nu_e/2\nu_g)^2$. We see the validity regime of MRWA depends
on $\eta$ for this case of single phonon JCM. Physically the trap
frequency difference can be controlled experimentally \cite{jeff}.
When $\nu_e=\nu_g$ as for an ion, we recover the condition $\eta
\ll 2$ as discussed by Moya-Cessa {\it et al.} \cite{moya}. Under
this condition, we arrive at the single phonon model
\begin{eqnarray}
\tilde{{\cal H}}^{\rm (1p)}_{MRWA}&=&{\cal H}_{\rm free}^{\rm (
1p)}+V^{\rm (1p)},
\nonumber\\
{\cal H}_{\rm free}^{(\rm 1p)}&=&\nu a^{\dagger}a+\frac{\nu}{2}\sigma_z, \nonumber\\
V^{\rm (1p)}&=&-ig(a^{\dagger}\sigma_- - a\sigma_+).
\label{eq:jcm}
\end{eqnarray}
This Hamiltonian is the same with the Jaynes-Cummings model
$H_{\rm JCM}=\nu a^{\dagger}a+(\nu/2)\sigma_z-gp\sigma_x$ with the
counter-rotating terms $a^\dag\sigma_+,a\sigma_-$ are dropped
under RWA.

Alternatively, there is a second resonance at $\Omega=2\nu$. The
condition for MRWA in this case is $\zeta/2 \ll \Omega$
\cite{peng}, which becomes $7\nu_e^2+9\nu_g^2\gg 0$. Hence, the
validity regime of MRWA becomes independent of $\eta$ in this case
and can be realized simply by choosing larger trap frequencies. In
contrast to the single phonon JCM in Eq. (\ref{eq:jcm}), the two
phonon squeezing model can be achieved beyond LDL and is described
by
\begin{eqnarray}
\tilde{{\cal H}}^{\rm (2p)}_{MRWA}&=&{\cal H}_{\rm free}^{\rm (2p)}
+V^{\rm (2p)},\nonumber\\
{\cal H}_{\rm free}^{\rm (2p)}&=&\nu a^{\dagger}a+\nu\sigma_z, \nonumber\\
V^{\rm (2p)}&=&\frac{\zeta}{2}(a^{\dagger 2}\sigma_-+a^2\sigma_+).
\label{eq:2p}
\end{eqnarray}

Equations Eq. (\ref{eq:jcm}) and Eq. (\ref{eq:2p}) consist the
major results of the general unitary transform introduced earlier.
Their complete dynamics, however, can be complicated as
transformation back into the original Schrodinger picture induces
coherent mixing of different states. In particular, if the system
is initially prepared in a state $\psi(0)$, the linearized model
Hamiltonian Eq. (\ref{eq:jcm}) has an initial state
$\tilde\psi(0)=T\psi(0)$, which is always a superposition of both
internal states. In other words, even for systems initially in
only one internal state such that it has a vanishing dipole
moment, the transformed system will always have a non-vanishing
dipole-moment. It is known in this case that counter rotating
terms (CRT) neglected in making MRWA become more significant and
validity regimes for MRWA more restricted \cite{seke}.

We are now in a position to discuss validity regimes of the
commonly used MRWA for the simplified model Hamiltonians Eqs.
(\ref{eq:jcm}) and (\ref{eq:2p}). We will
focus on the single-phonon model in Eq. (\ref{eq:jcm}) as it
can be compared directly to existing results (under MRWA) in
literature \cite{moya}. Furthermore, while the single phonon JCM
can be obtained for small restricted values of the Lamb-Dicke
parameter, two-phonon model is obtained without any restriction
on the Lamb-Dicke parameter. Thus, determining the validity regime of
MRWA is more essential for the single-phonon JCM.

\section{Numerical method}
In the following discussion we focus on the single-phonon model
and thus ignore the unnecessary superscript ${\rm (1p)}$ for
notational simplicity. In the numerical studies, for the
approximate JCM Eq. (\ref{eq:jcm}) we propagate any given initial
states according to ${\tilde\psi}(t)=e^{-i{\cal H}_{\rm free} t}
U_I(t)\tilde{\psi}(0)$ with the known propagator $U_I(t)=e^{-itV}$
in the interaction picture \cite{scully},
\begin{eqnarray}
U_I(t)=\frac{1}{\sqrt{2}} \left (
\begin{array}{c c}
\cos[{gt\sqrt{aa^\dag}}] & \sin[{gt\sqrt{aa^\dag}}]
\frac a {\sqrt{aa^\dag}}\\
-{a^\dag\over \sqrt{aa^\dag}}{\sin[{gt\sqrt{aa^\dag}}]} &
\cos[{gt\sqrt{a^\dag a}}]
\end{array} \right ).
\end{eqnarray}
Transformed states are denoted by
$\tilde{\psi}=\tilde{T}\psi$. In the Schrodinger picture, where
observables are computed, the wave function becomes
$\psi(t)=\tilde{T}^\dag e^{-i{\cal H}_{\rm
free}t}U_I(t)\tilde{T}\psi(0)$. For example, the mean phonon
number $\langle a^\dag a\rangle$ is found to be
\begin{eqnarray}
\langle n\rangle=\psi(0)^\dag \tilde{T}^\dag U_I^\dag e^{i{\cal
H}_{\rm free}t}\tilde{T}a^\dag a \tilde{T}^\dag e^{-i{\cal H}_{\rm
free} t}U_I(t)\tilde{T}\psi(0),
\end{eqnarray}
which can be further simplified with the use of
displacement operator property
\begin{eqnarray}
\tilde{T}a^{\dagger}a\tilde{T}^\dag
=(a^{\dagger}a+\frac{\eta^2}{2})\hat{I}-i\frac{\eta}{2}
(a^{\dagger}-a)\sigma_{x}.
\end{eqnarray}
We note that the last term in the above Eq. involves
momentum operator. Therefore, it does not commute with
${\cal H}_0$ and introduces fast oscillations due to
the $e^{-i2\nu t}$ factor. We finally get,
\begin{eqnarray}\label{eq:n}
\langle n\rangle=\tilde{\psi}(0)^{\dagger}U_I^{\dagger}\left[
(a^{\dagger}a+\frac{\eta^2}{2})\hat{I}-i\frac{\eta}{2}
(a^{\dagger}e^{i\nu t}-ae^{-i\nu t})(\sigma_+e^{i\Omega t}
+\sigma_-e^{-i\Omega t})\right]U_I \tilde{\psi}(0).
\end{eqnarray}
We immediately see that CRT contribute to the
dynamics of $\langle n\rangle$ because of
multiple transitions in the original Hamiltonian.
In addition to $\langle n\rangle$, CRT also affect other
dynamical variables of interest, e.g. the Mandel-Q factor
$(\langle (a^\dag a)^2\rangle-\langle a^\dag a\rangle^2)/2$,
which characterizes phonon statistics and the
impurity factor ${\cal I}=1-{\rm tr}(\rho_{p}^2)$\cite{julio}.
$\rho_p$ is the reduced density operator in the phonon subsystem.

When all CRT are included, i.e. no MRWA is made, it is no longer
possible to solve dynamic propagation analytically. However, the
transformed Hamiltonian can be diagonalized numerically in a
truncated phonon Fock state basis. It is easy to check the
accuracy of such a truncation by testing for convergence with
successively larger basis states. For an initial Poisson
distribution of vibrational phonons, the dimension of the truncated
space becomes small enough that efficient algorithms are
readily available. To illustrate this, let us consider an initial
state where the particle is in the internal state $|e\rangle$ with
a motional coherent state $|\alpha\rangle$, i.e.
$\psi(0)=|e\rangle|\alpha\rangle$. The transformed initial
condition becomes
\begin{eqnarray}
\tilde{\psi}(0)=e^{-i\eta \Re(\alpha)}\frac{|e\rangle-|g\rangle}{\sqrt{2}}
|\alpha-i\frac{\eta}{2}\rangle.
\label{psii}
\end{eqnarray}
In a truncated Fock space of dimension $N$ ($\gg |\alpha-i\eta/2|^2$),
the state vector is expanded as
\begin{eqnarray}
\tilde{\psi}(t)=\sum_{n=1,\Lambda=e,g}^{N}X_{n+n_{\Lambda}}(t)|\Lambda n\rangle,
\end{eqnarray}
with $n_{\Lambda}=0,N$ and $\Lambda=e,g$ respectively.
The initial condition (\ref{psii}) then becomes
\begin{eqnarray}
X_{n+n_{\Lambda}}(0)=s_{\Lambda}\frac{1}{\sqrt{2}}e^{-i\eta Re(\alpha)}
F_{n-1}(\alpha-i\frac{\eta}{2}),
\end{eqnarray}
with $s_{\Lambda}=+1,-1$. The coherent
state probability amplitudes in Fock space are
$F_n(\alpha)=\exp{(-|\alpha|^2/2)}\alpha^n/\sqrt{n!}$.
After back transformation, the original state vector evolution
is then determined by
\begin{eqnarray}
\psi(t)=\sum_{n=1,\Lambda=e,g}A_{n+n_{\Lambda}}(t)|\Lambda n\rangle,
\end{eqnarray}
with coefficients $A_{j}$ for $(j=1,2,\cdots,2N)$ given by
\begin{eqnarray}
A_{n+n_{\Lambda}}(t)=\frac{1}{\sqrt{2}}\sum_{m=1}^{N}
D_{n-1,m-1}(s_{\Lambda}\frac{\eta}{2})[X_{m}(t)+s_{\Lambda}X_{m+N}(t)].
\end{eqnarray}
The displacement operator matrix elements in Fock basis are known
analytically in terms of Laguerre polynomials \cite{wong}.
Therefore, the complete time evolution is obtained provided the
transformed Hamiltonian is diagonalized. This is achieved through
the standard expression
\begin{eqnarray}
X_{i=1,...,N}(t)=\sum_{j,k=1}^{2N}V_{ij}V^{-1}_{jk}e^{-iW_{j}t}X_{k}(0),
\end{eqnarray}
where $W_{j}$ are the eigenvalues of the transformed Hamiltonian
and $V$ is the diagonalizing matrix whose columns are the
corresponding eigenvectors. In the following section we will
compare MRWA results with the numerical diagonalization method
(NDM).
\section{Results and discussions}
\label{sec:res}
\begin{figure}
\centerline{\epsfig{file=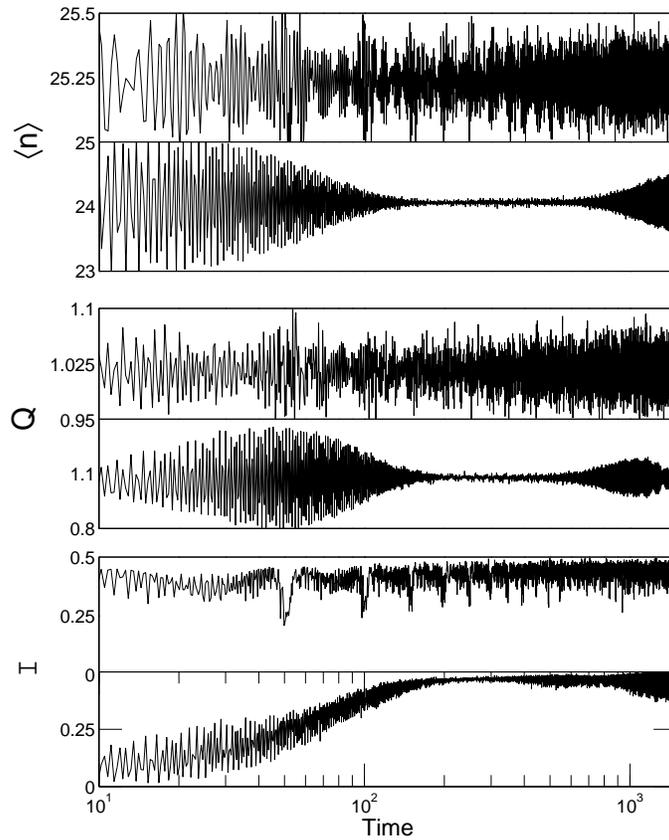, width=9.cm}}
\caption{Comparison of the mean phonon number $\langle n\rangle$,
the Mandel-Q factor, and the impurity parameter ${\cal I}$ for
$\eta=0.5$ and $\alpha=(0.5,5)$. Always, the upper sub-figures are
from the exact numerical diagonalization, while the lower ones are
from MRWA. Time axis is in dimensionless from (scaled by $g=\eta\nu/2$).
Note the differences as compared with Ref. [3].}
\label{fig1}
\end{figure}

\begin{figure}
\centerline{\epsfig{file=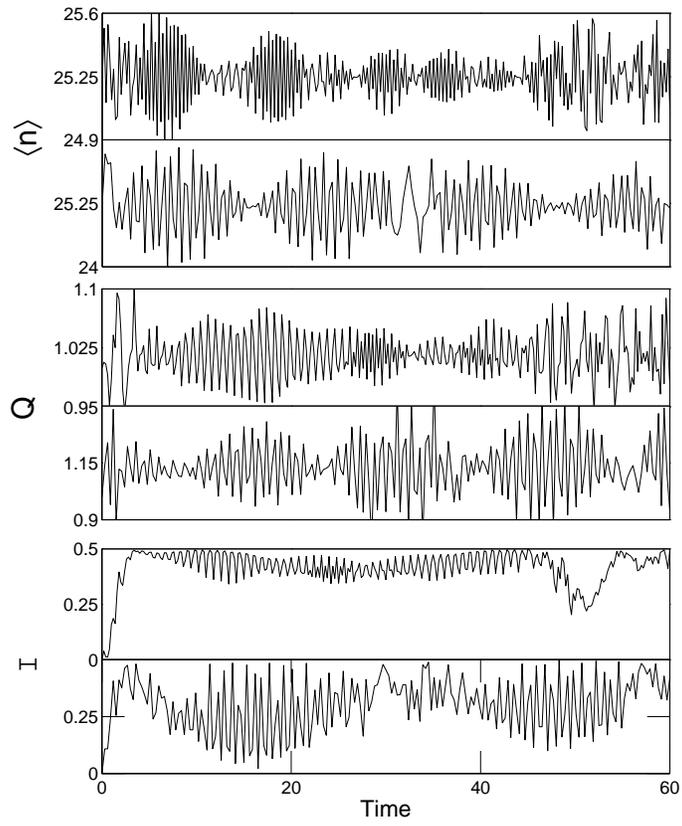, width=9.cm}}
\caption{The same as in Fig. 1 but for
$\eta=0.5$ and $\alpha=(5,0.5)$. }
\label{fig2}
\end{figure}

\begin{figure}
\centerline{\epsfig{file=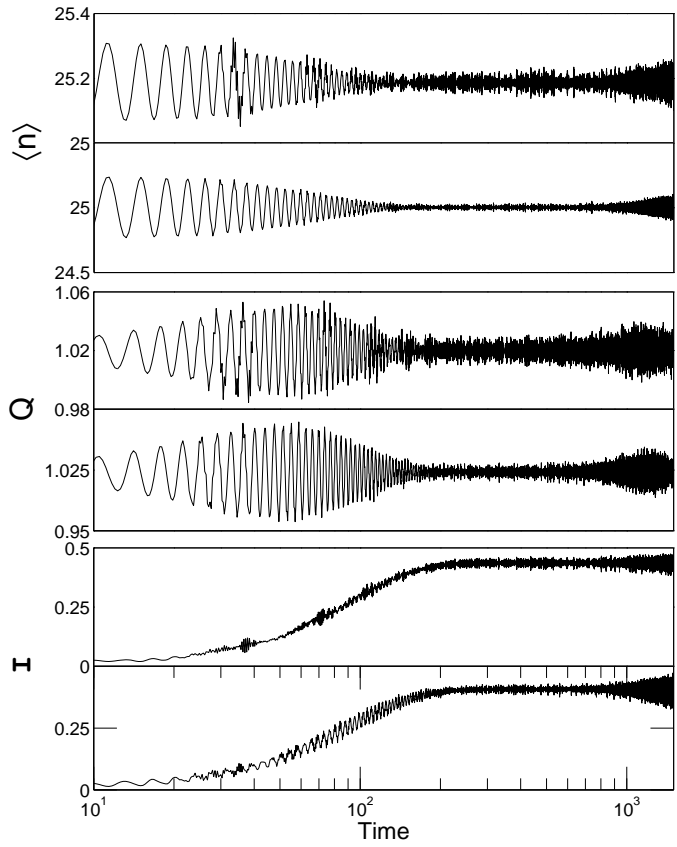, width=9.cm}}
\caption{The same as in Fig. 1 but for
$\eta=0.1$ and $\alpha=(0.5,5)$. Note the super-revivals for $Q$.}
\label{fig3}
\end{figure}

\begin{figure}
\centerline{\epsfig{file=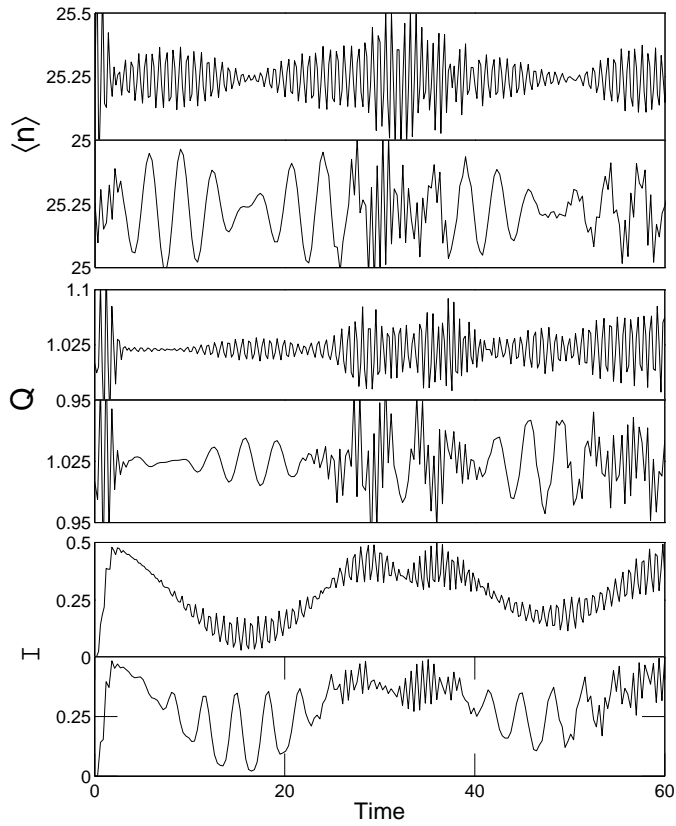, width=9.cm}}
\caption{The same as in Fig. 1 but
for $\eta=0.1$ and $\alpha=(5,0.5)$.}
\label{fig4}
\end{figure}

We focus in this section on the comparison of exact numerical
results with results from MRWA for an initial motional
coherent state $|\alpha\rangle$ with
$\alpha=|\alpha|\exp{(i\beta)}$. Moya-Cessa {\it et al.}
\cite{moya} have discussed a restricted form of the unitary
transformation that led to a similar linearized model
Hamiltonian earlier. Although formally independent of the Lamb
Dicke parameter, their results were obtained under MRWA. In
this first example, we demonstrate that their illustrative figure
as presented in Ref. \cite{moya} is in fact invalid at the
presumed marginal LDL $\eta=0.5$. In Fig. \ref{fig1}, we see that
MRWA result predicts regular behaviors for several dynamical
variables, while the exact result from the NDM at $\eta=0.5$ and
$\alpha=(0.5,5)$ display significant differences. In general we
find that with NDM, the initial state losses its purity faster and
ends up with a larger value for the time averaged mean phonon
number. On the other hand, MRWA results typically give
larger widths of temporal fluctuations for the Mandel-Q factor,
especially at earlier times. Perhaps most importantly, MRWA
results predict super-revivals \cite{moya} in $Q$ and $\langle n\rangle$
that were never observed with NDM.
Differences of similar orders of magnitude are also found
for $\alpha=(5,0.5)$ as detailed in Fig. \ref{fig2}.
We also note the significantly
improved quantitative agreement for $\langle n\rangle$ in this
case. In fact, such improved agreement with MRWA results
always seem to occur when $\beta\approx 0$.
Now we compare the deeper LDL regime of $\eta=0.1$. As shown in Figs.
\ref{fig3} and \ref{fig4}, noticeable errors with
MRWA were still found, although at significantly reduced levels
as compared with Figs. \ref{fig1} and \ref{fig2}.
Figure \ref{fig4} compares early time dynamics for the same
$\eta=0.1$ but with $\alpha=(5,0.5)$. We note that the ${\cal
I}=0$ value predicts the existence of Schrodinger cat states from
both MRWA and NDM results \cite{julio}. Also the agreement for
$\langle n\rangle$ and $Q$ are better than the case of
$\alpha=(0.5,5)$. These results show clearly that errors in
MRWA are sensitive to the phase of the initial coherent state
amplitude.

\begin{figure}
\centerline{\epsfig{file=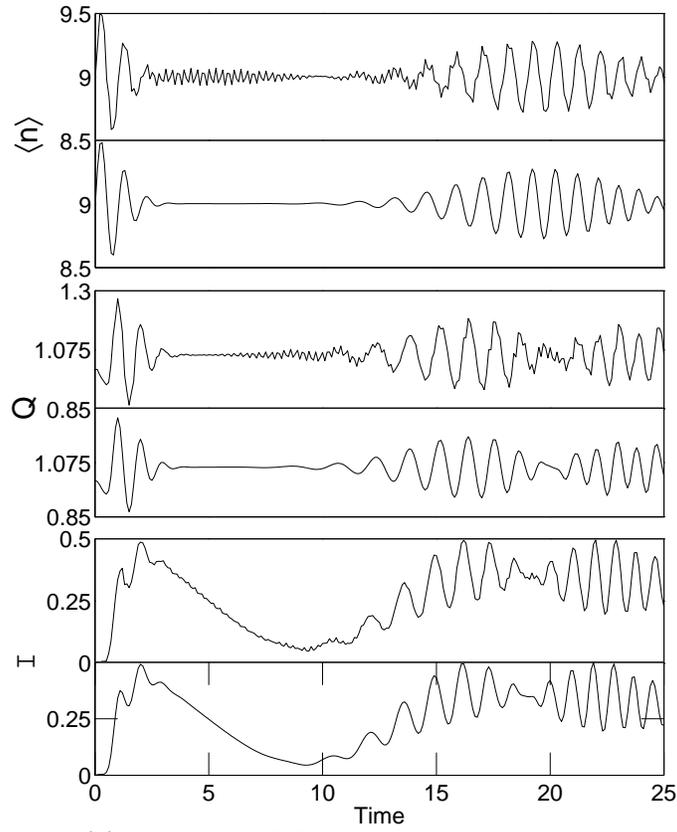, width=9.cm}} \caption{The mean
phonon number $\langle n\rangle$, the Mandel-Q factor, and the
impurity parameter ${\cal I}$ with $\eta=0.1$ and $\alpha=(3,0)$
for the simple single-phonon JCM $H_{\rm JCM}$ with and without
RWA. In both cases no back transformation is employed but an
initial condition of the form $\tilde{T}\psi(0)$ has been used to
compare with the multiple-phonon transition effects in
Fig. \ref{fig6}. Within each sub-figure, the upper (lower) part is
for $H_{\rm JCM}$ with (without) counter rotating terms.}
\label{fig5}
\end{figure}

\begin{figure}
\centerline{\epsfig{file=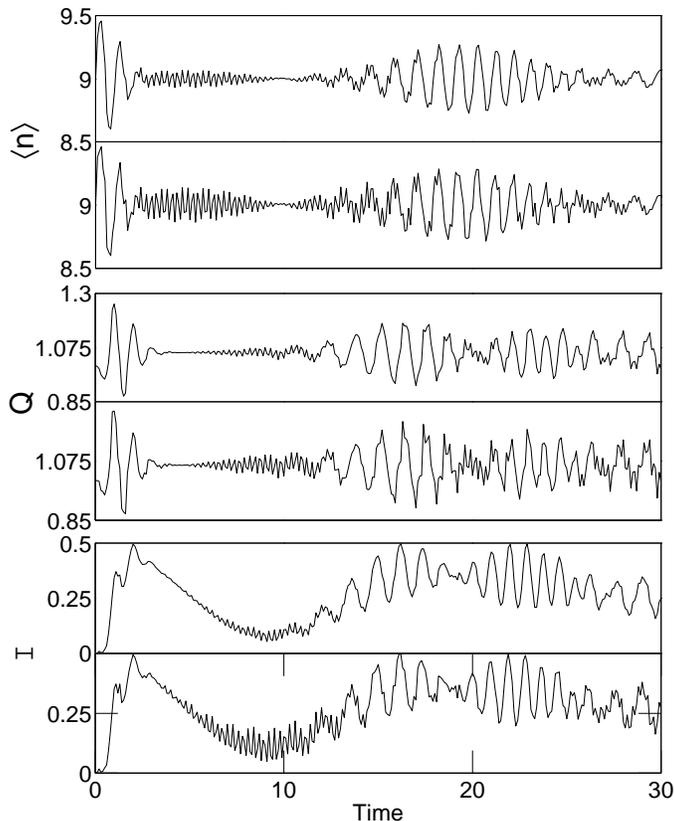, width=9.cm}} \caption{The same
as in Fig. 5 but now for the full ion-trap Hamiltonian which
includes multiple phonon transition effects. Within each
sub-figure, the upper (lower) curve is obtained by NDM (MRWA).}
\label{fig6}
\end{figure}

To further elucidate effects of MRWA, we have carried out
additional comparisons. In Fig. \ref{fig2} and in Fig. \ref{fig4},
we notice that MRWA results predict larger oscillations than the
actual smoother behavior of $\langle n\rangle$, $Q$, and $I$.
This observation is in contrast to our expectations that stem from
the effects of RWA in the usual JCM. Let us first recall what is
the effect of neglecting counter rotating terms in the $H_{JCM}$.
For comparation, we consider an initial condition of the
form $\tilde T\psi(0)$. The results obtained by propagating
$H_{JCM}$ dynamically with and without counter-rotating terms are
given in Fig. \ref{fig5}. We see $H_{JCM}$ with RWA
predicts smoother behavior while the actual $H_{JCM}$ results
carry small nutations due to counter-rotating effects.
For the ion-trap system displayed in Fig. \ref{fig6},
the results are just the opposite. In ion-trap
Hamiltonian MRWA results carry more
nutations than the actual smoother behavior. The basic
difference between the JCM and the ion-trap system is the
requirement of the additional back transformation for the latter via
$\tilde{T}^{\dagger}$ in determining the wave function
evolution. From the physical point of view, this transformation
brings back the effect of multiple-phonon transitions,
present in the ion-trap case, after a simplified single phonon
transition dynamics has been determined conveniently in the
interaction picture. Neglecting
counter-rotating terms in the effective single-phonon model Eq. (\ref{eq:jcm}),
however, ignores too many multiple-phonon processes at the end
of the transformation,
thus introduce more noise around the otherwise smoother
collapse-revival patterns. We note that this (MRWA in ion-trap models)
is opposite to effects of the RWA in the JCM and has been
found for an initial motional state with almost real coherent
state amplitudes, i.e. $\beta\approx 0$.

\section{conclusion}
\label{sec:conc}
We have investigated the regimes of validity for MRWA of a trapped
particle under its coherent interaction with a plane wave laser
field. Our study is based on a family of general unitary
transformations that incorporate several earlier transforms as
special cases \cite{moya,you}. This general transformation
facilitates the linearization of the system Hamiltonian including
the particle's motional degrees of freedom. In model studies
presented here, two distinct dynamical regimes appear, the
single-phonon dynamics and the double-phonon dynamics.
The single phonon model is standard in the deep LDL.
The two-phonon model, on the other hand, does not impose any
restrictions on the Lamb-Dicke parameter. Thus, addressing
validity regimes of MRWA is only necessary for
the single-phonon dynamics used extensively in the literature.
For this aim, we have employed a numerical diagonalization
procedure (without MRWA) and comparatively assessed the validity
conditions of MRWA. We find that remarkable errors exist in
regimes where MRWA was believed to be applicable. Furthermore, our
simulations show that the accuracy and detailed quantitative
effects of MRWA depends on the
phase of the initial motional coherent state. When this phase is
close to zero, time averaged mean phonon number $\langle n\rangle$
displays improved agreement with MRWA results. The agreement
however is far from perfect even in the deep LDL. Qualitatively,
for almost real initial coherent state amplitude, the counter
rotating terms in ion-trap model lead to smoother dynamical
behaviors than the results obtained under MRWA, which introduces
more larger nutations around the actual more smooth dynamical
patterns. This is in stark contrast to CRT effects in usual JCM
problems. Physically, this effect arises because of the existence
of multiple transitions in motional states. Our study indicates
that the connection between cavity QED and ion-trap systems in
JCM-type formalisms are possible as long as ranges of
the Lamb-Dicke parameter
are carefully analyzed to enforce the validity of MRWA.

\section{ACKNOWLEDGMENT}
This work is supported by a grant from
the National Security Agency (NSA), Advanced Research and
Development Activity (ARDA), and the Defense
Advanced Research Projects Agency (DARPA) under Army Research Office
(ARO) Contract No. DAAD19-01-1-0667 and by the NSF grant No. PHY-9722410.




\end{document}